\documentclass[journal]{IEEEtran}
\usepackage{cite}

\usepackage{color,soul}
\usepackage{gensymb}
\usepackage{dsfont}

\ifCLASSINFOpdf
  \usepackage[pdftex]{graphicx}
  \graphicspath{{../pdf/}{../jpeg/}}
  \DeclareGraphicsExtensions{.pdf,.jpeg,.png}
\else
  \usepackage[dvips]{graphicx}
  \graphicspath{{../eps/}}
  \DeclareGraphicsExtensions{.eps}
\fi
\usepackage[cmex10]{amsmath}
\usepackage{amssymb}

\DeclareMathOperator*{\argminA}{arg\,min}
\usepackage{euscript}

\DeclareMathAlphabet{\mathpzc}{OT1}{pzc}{m}{it}
\usepackage{multirow}
\usepackage{algorithm}
\usepackage[noend]{algpseudocode}
\usepackage{diagbox}
\usepackage{physics}
\usepackage{tikz}

\ifCLASSOPTIONcompsoc
  \usepackage[caption=false,font=normalsize,labelfont=sf,textfont=sf]{subfig}
\else
  \usepackage[caption=false,font=footnotesize]{subfig}
\fi

\usepackage{url}

\begin{document}
%
\title{Mitigating Smart Meter Asynchrony Error Via Multi-Objective Low Rank Matrix Recovery
}
%
%
%

\author {Yuxuan~Yuan,~\IEEEmembership{Graduate Student Member,~IEEE,}
Kaveh~Dehghanpour,
	and Zhaoyu~Wang,~\IEEEmembership{Senior Member,~IEEE}
\thanks{This work was supported by the Advanced Grid Modeling Program at the U.S. Department of Energy Office of Electricity under Grant DE-OE0000875, and  the National Science Foundation under ECCS 1929975. (\textit{Corresponding author: Zhaoyu Wang})

Y. Yuan, K. Dehghanpour, and Z. Wang are with the Department of
Electrical and Computer Engineering, Iowa State University, Ames,
IA 50011 USA (e-mail: kavehdeh1@gmail.com; wzy@iastate.edu).

}
}
%
%

\markboth{Submitted to IEEE for possible publication. Copyright may be transferred without notice}%
{Shell \MakeLowercase{\textit{et al.}}: Bare Demo of IEEEtran.cls for Journals}
%



\maketitle

\begin{abstract}
Smart meters (SMs) are being widely deployed by distribution utilities across the U.S. Despite their benefits in real-time monitoring. SMs suffer from certain data quality issues; specifically, unlike phasor measurement units (PMUs) that use GPS for data synchronization, SMs are not perfectly synchronized. The asynchrony error can degrade the monitoring accuracy in distribution network. To address this challenge, we propose a  principal component pursuit (PCP)-based data recovery strategy. Since asynchrony results in a loss of temporal correlation among SMs, the key idea in our solution is to leverage a PCP-based low rank matrix recovery technique to maximize the temporal correlation between multiple data streams obtained from SMs. Further, our approach has a novel multi-objective structure, which allows utilities to precisely refine and recover all SM-measured variables, including voltage and power measurements, while incorporating their inherent dependencies through power flow equations. We have performed numerical experiments using real SM data to demonstrate the effectiveness of the proposed strategy in mitigating the impact of SM asynchrony on distribution grid monitoring.
\end{abstract}

\begin{IEEEkeywords}
Smart meters; sensor asynchrony; low rank matrix recovery; multi-objective optimization
\end{IEEEkeywords}

\section*{Nomenclature}
\addcontentsline{toc}{section}{Nomenclature}
\begin{IEEEdescription}[\IEEEusemathlabelsep\IEEEsetlabelwidth{$V_1,V_2,V_3$}]
\item[BCSE] Branch current state estimation
\item[DSSE] Distribution system state estimation
\item[MPE] Mean percentage error
\item[PCP] Principle component pursuit
\item[PCA] Principle component analysis
\item[SM] Smart meter
\item[WLS] Weighted least squares
\item[$G$] Gain matrix
\item[$H$] Jacobian matrix
\item[$h_i$] Measurement function that maps state values to the measurement variable $i$
\item[$\pmb{I_{re}},\pmb{I_{im}}$] Current real and imaginary values for all the branches
\item[$J$] Sum of squared residuals
\item[$L$] Weight parameter for penalizing deviations from SM measurements
\item[$M_U$] Voltage observation matrix
\item[$M_U^*$] Refined post-mitigation voltage matrix
\item[$M_P$] Nodal active power injection matrix
\item[$M_P^*$] Refined post-mitigation active power matrix
\item[$M_Q$] Nodal reactive power injection matrix
\item[$M_Q^*$] Refined post-mitigation reactive power matrix
\item[$M_{MV}$] Synchronized sensor measurements
\item[$M_{z}$] Measurement vector
\item[$M_{PS}$] Pseudo measurements
\item[$P_i(t_j)$] Measured active power at node $i$ at time $t_j$
\item[$Q_i(t_j)$] Measured reactive power at node $i$ at time $t_j$
\item[$R$] Branch resistance matrix of the system
\item[$U_0$] Squared voltage magnitude of substation
\item[$U_i(t_j)$] Measured voltage magnitude squared at node $i$ at time $t_j$
\item[$W$] Weight matrix
\item[$X$] Branch reactance matrix of the system
\item[$\pmb{x_s}$] System state vector
\item[$Y_M, Y_S$] Interim matrices using the latest solution updates
\item[$Z_M, Z_S$] Interim matrices using the full history of the solution trajectory
\item[$\alpha,\beta$] Auxiliary matrices
\item[$\Delta S_U$] Asynchrony voltage error matrix
\item[$\Delta E_U$] Voltage measurement error matrix
\item[$\Delta S_P$] Asynchrony active power error matrix
\item[$\Delta E_P$] Active power measurement error matrix
\item[$\Delta S_Q$] Asynchrony reactive power error matrix
\item[$\Delta E_Q$] Reactive power measurement error matrix
\item[$\delta_U, \delta_P, \delta_Q$] Standard deviations of voltage, active power, and reactive power measurement errors
\item[$\Gamma(\cdot,\cdot)$] Differentiable function for low rank matrices
\item[$\Gamma_T$] Total approximate sparsity norm for all the SM datasets
\item[$||\cdot||_{*}$] Nuclear norm operation
\item[$||\cdot||_{1}$] 1-norm operation
\item[$||\cdot||_{F}$] Frobenius norm operation
\item[$<\cdot,\cdot>$] Frobenius inner product
\item[$\lambda_U, \lambda_P, \lambda_Q$] Balanced parameters for voltage, active power, and reactive power measurements
\item[$\mu_U,\nu_U$] Smoothness parameter
\item[$\omega_1, \omega_2, \omega_3$] Non-negative weights
\item[$\Psi(\cdot,\cdot)$] Differentiable function for sparse error matrices
\item[$\sigma_i^2$] Error variance of sensor $i$
\item[$\tau(\cdot,\cdot)$] Aggregate gradient factor 
\item[$\zeta_j(A)$] $j$'th singular value of an arbitrary matrix $A$
\end{IEEEdescription}

%
\IEEEpeerreviewmaketitle

\section{Introduction}
The wide-scale deployment of smart meters (SMs) provides a unique opportunity for utilities to enhance their situational awareness capabilities in distribution grids. By 2018, more than 150 million customers across the U.S. were equipped with SMs \cite{EIA}. On the other hand, SMs are commonly counted among low-quality sensors. Specifically, SMs are asynchronous due to mismatching in sampling time among sensors in the grid, which can limit their applicability in real-time system monitoring \cite{Lin2019}.

Most previous works on distribution grid state estimation have assumed that SMs are perfectly time-aligned \cite{Primadianto2017,Deh2018}. Only few works have studied the impact of time misalignment and asynchrony of various sensors on grid monitoring and situational awareness: In \cite{Antonios2019,Schwefel2018}, the statistical characteristics of time misalignment in distribution grid sensors have been estimated using Markov-modulated models. In \cite{Lin2019}, exponential load variation trends are exploited for developing confidence intervals for SM data samples in distribution system state estimation (DSSE) to compensate for time delays and asynchrony. In \cite{Carvaro2019}, a dynamic DSSE formulation is proposed for multitude of asynchronous sensors, which has proven bounded estimation errors. In \cite{Bolognani2015,Alimardani2015}, meter clock synchronization errors are captured through Gaussian probability distributions and represented in DSSE. This idea was also applied in \cite{Ni2016} to model measurement errors in grid monitoring. Most solutions proposed for mitigating SM data quality issues rely on \textit{a priori} knowledge of error distribution structure and parameters, which can be difficult to acquire due to information scarcity. 

In this paper, we propose a SM data recovery technique that is capable of mitigating the impact of asynchrony error in grid monitoring. Our method has three novel features: (1) We have noted that a rise in SM asynchrony results in a loss of mutual temporal correlation in their time-series data streams. This loss of temporal correlation can be translated into an increase in the rank of \textit{observation matrices}, which store the measurement data from multiple SMs. Thus, we propose to cast the asynchrony error mitigation problem as a low rank matrix recovery process. For this purpose, we have leveraged principle component pursuit (PCP) techniques \cite{Bouwmans2014,Rodriguez2013}. PCP employs data-centric optimization for decomposing SM datasets to identify and separate asynchrony error term from raw data. The main idea is that by manipulating the SM data and reducing the rank of the observation matrices, we will enhance the temporal correlation among the SMs which rolls back the adverse impact of asynchrony. (2) In addition to asynchronous errors, SM data has measurement errors that result from the imprecision (i.e., noise) of the measuring devices. Typically, SMs have a relative measurement error of about 1$\%$. Further, unlike image datasets, synchronous SM measurements and asynchronous errors cannot be exactly low rank and exactly sparse. These data properties hinder the applications of state-of-the-art low rank data recovery methods to deal with SM asynchrony errors, such as robust principal component analysis (PCA) \cite{Zhou2010}. To deal with these problems, we utilize a relaxation to PCP that introduces an entry-wise noise term to represent SM measurement errors in the objective function and eliminate rank-1 constraints. (3) SMs are multi-modal, meaning that they can measure several different variables, including nodal voltage magnitude and nodal average active power (plus nodal reactive power, in some cases.) To mitigate the impact of sensor asynchrony, data recovery needs to be conducted over all measurement datasets simultaneously. However, since these multi-modal datasets are inherently interdependent due to the grid physics, a coordination scheme is required to revise all the datasets while capturing their dependencies. To achieve this, we propose a new multi-objective data recovery formulation that refines voltage magnitude, active/reactive power measurements (and pseudo-measurements), concurrently. The dependencies among these datasets are captured via approximate DistFlow-based constraints \cite{Gilbert1998,Baran1989}. We have developed a Nesterov-based technique to solve the PCP-based multi-objective optimization for recovering multiple SM datasets \cite{Aybat2011}.

The main contributions of this paper are summarized as follows: 
\begin{itemize}
\item An important observation from real data is presented: asynchrony results in loss of temporal correlation among neighboring SMs. This observation can be quantified using the rank of the nodal voltage observation matrix.
\item A novel low rank-based data recovery method is developed to fully mitigate asynchronization error in grid monitoring based on our observation.
\item The proposed method considers various specific properties of SM data for enhancing the quality of the recovered data and ensure consistency with grid physics: 1) SMs can measure several different asynchronous variables; 2) SM measurements are statistically interdependent; 3) small entry-wise measurement errors exist within SM measurements. 
\item Our method handles SM asynchrony issue without needing high-resolution reference sensors, such as micro-PMUs, which are unavailable in most practical distribution systems.
\item The proposed solution has been tested using real SM data and feeder models to verify its performance. 
\end{itemize}


The rest of the paper is constructed as follows: Section \ref{sec:MO} presents the proposed multi-objective data recovery method and our approximate first-order solution; Section \ref{sec:dsse} demonstrates the application of data recovery in grid monitoring; Section \ref{sec:result} analyzes numerical results and verification of the proposed models; finally, Section \ref{sec:con} presents the paper conclusions.

\section{Multi-Objective SM Data Recovery Strategy}\label{sec:MO}
In this section, we lay out our data recovery solution for mitigating the errors caused by the asynchronous nature of SMs in distribution grids. This includes key ideas in developing a multi-objective optimization formulation, along with an approximate first-order algorithm to solve the model.
\subsection{Rationale}
The available data from SMs can be organized into \textit{observation matrices}. These matrices capture the time-series measurements of several sensors within a given time window $[t_1,t_m]$. For example, the voltage observation matrix is as follows:  
\begin{equation}
\label{eq:MU}
M_U = \left[
\begin{array}{ccc}
U_1(t_1) &  \cdots & U_N(t_1)\\
\vdots &  \ddots & \vdots\\
U_1(t_m) &  \cdots & U_N(t_m)
\end{array}
\right]
\end{equation}
where, $U_i(t_j)$ is the measured voltage magnitude squared at node $i$ and at time $t_j$. Note that each column of $M_U$ corresponds to an SM. The observation matrices can be constructed at feeder-, lateral-, or service transformer-levels.
\begin{figure}
\centering
\includegraphics[width=1\columnwidth]{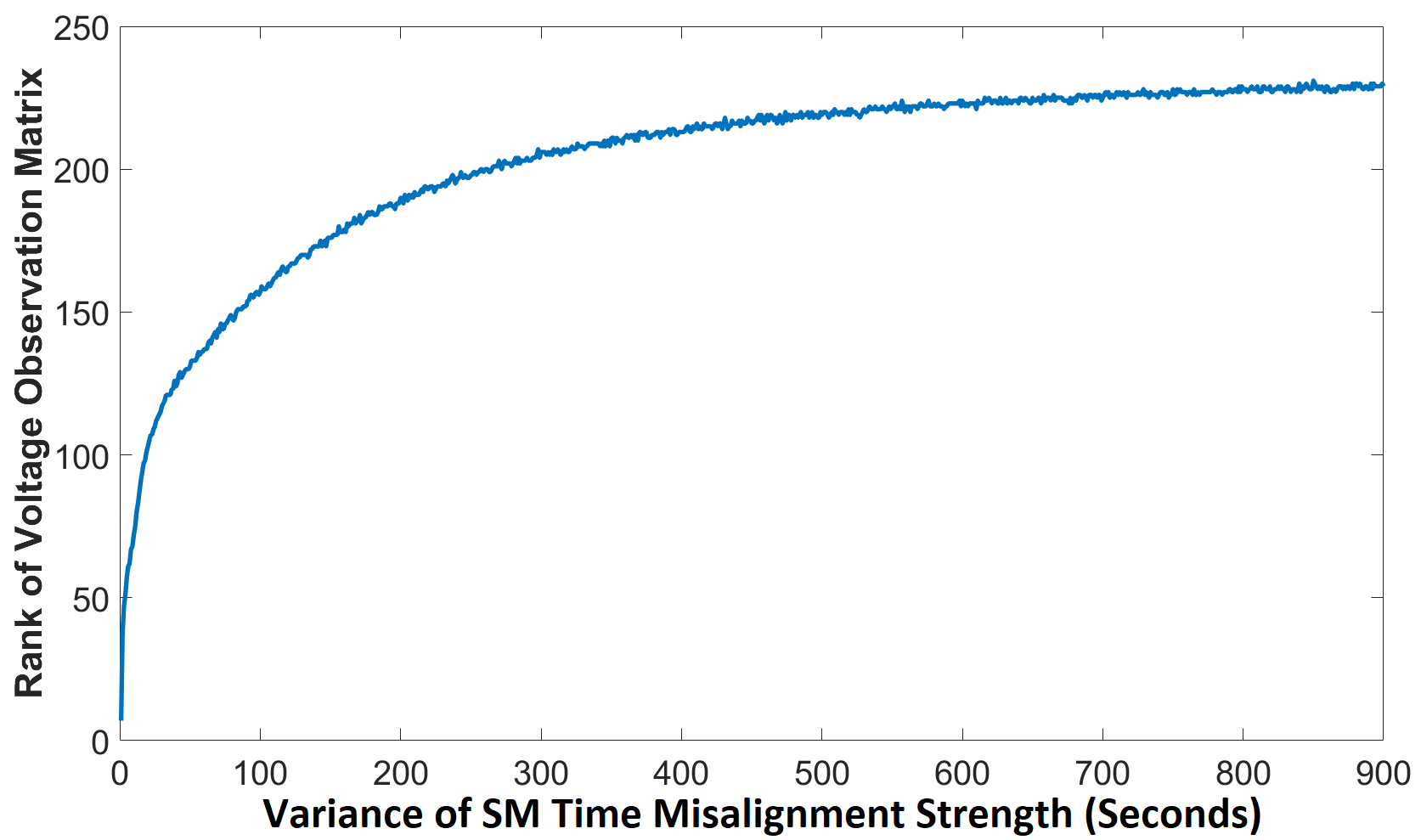}
\caption{Rank increase in $M_U$ due to SM asynchrony.}
\label{fig:rank}
\end{figure}

Our PCP-based data recovery model is based on a key observation from real data: asynchrony among SMs leads to an increase in the rank of $M_U$. The increase in rank is caused by loss of temporal correlation among SMs, which translates into a decrease in statistical correlations in columns of $M_U$ (i.e., the columns lose linear dependency.) This observation can be backed-up by numerical experiments, as shown in Fig. \ref{fig:rank}. This figure shows the average rank of $M_U$ at various time windows (measured for a grid lateral) as a function of strength of SM asynchrony (measured in terms of variance of time misalignment distribution.) As is observed, the rank of the observation matrix increases as the SM asynchrony intensifies. Note that this observation can be found on the data from SMs with diverse resolutions, including 15 minutes, 30 minutes, and 60 minutes.

To fully capture and mitigate the impact of SM asynchrony, similar observation matrices can be defined for nodal active and reactive power injection measurements, denoted as $M_P$ and $M_Q$, respectively:
\begin{equation}
\label{eq:MP}
M_P = \left[
\begin{array}{ccc}
P_1(t_1) &  \cdots & P_N(t_1)\\
\vdots &  \ddots & \vdots\\
P_1(t_m) &  \cdots & P_N(t_m)
\end{array}
\right]
\end{equation}
\begin{equation}
\label{eq:MQ}
M_Q = \left[
\begin{array}{ccc}
Q_1(t_1) &  \cdots & Q_N(t_1)\\
\vdots &  \ddots & \vdots\\
Q_1(t_m) &  \cdots & Q_N(t_m)
\end{array}
\right]
\end{equation}
where, $P_i(t_j)$ and $Q_i(t_j)$ are active and reactive power measurements at node $i$ and time $t_j$, respectively. Note that in general SMs are capable of measuring both average active and reactive powers. However, in many cases, this function is not activated for residential sensors. Thus, in case the reactive power data is unavailable, \textit{pseudo-measurements} can be applied instead to construct an approximate $M_Q$. Note that our method is robust to gross sparse errors, thus, it can handle the uncertainty of pseudo-measurements and low quality data.

\subsection{Data Recovery Model}
The main component of asynchrony error mitigation is to compensate for the loss of temporal correlation among SMs. Since this loss can be detected via the changes in the ranks of the observation matrices, asynchrony error mitigation can be written as a low rank matrix recovery model. To consider both asynchrony errors and small entry-wise measurement errors in SM data, our data recovery approach models an observation matrix (i.e., asynchrony voltage magnitude matrix) as the summation of three components: a low rank voltage magnitude matrix, an asynchrony error matrix, and a measurement error matrix. The goal is to identify unknown voltage magnitude matrix and asynchrony error matrix  within the datasets in the presence of entry-wise noise. The model is shown below:
\begin{equation}
\label{eq:modelU}
M_U = M_U^* + \Delta S_U + \Delta E_U
\end{equation}
where, $M_U^*$ represents the \textit{refined post-mitigation} voltage magnitude matrix which has a low rank, $\Delta S_U$ is the asynchrony error matrix, and $\Delta E_U$ represents entry-wise measurement errors. It should be noted that measurement error is different from asynchrony error, as mentioned in previous work \cite{Alimardani2015}. The same representation applies to both active and reactive measurements and pseudo-measurements, as follows:
\begin{equation}
\label{eq:modelP}
M_P = M_P^* + \Delta S_P + \Delta E_P
\end{equation}
\begin{equation}
\label{eq:modelQ}
M_Q = M_Q^* + \Delta S_Q + \Delta E_Q
\end{equation}
where, the sub-components are defined similar to \eqref{eq:modelU}. The objective of the data recovery process is to revise the SM data in a way that the ranks of observation matrices are minimized (i.e., temporal correlations among SMs are maximized), while the extent of changes made in the original data is kept at a minimum level. This goal can be represented using three objective functions, corresponding to the available datasets, $M_U$, $M_P$, and $M_Q$, as follows:
\begin{equation}
\label{eq:all_obj}
\begin{cases}
f_U = ||M_U^*||_{*} + \lambda_U||\Delta S_U||_1\\
f_P = ||M_P^*||_{*} + \lambda_P||\Delta S_P||_1\\
f_Q = ||M_Q^*||_{*} + \lambda_Q||\Delta S_Q||_1
\end{cases}
\end{equation}
where, $||\cdot||_{*}$ and $||\cdot||_1$ are the nuclear norm and 1-norm (i.e., sparsity norm) operations, respectively. These norms are calculated as follows \cite{Phillips1996}:
\begin{equation}
\label{eq:nuc}
||A||_{*} = \sum_j\zeta_j(A)
\end{equation}
\begin{equation}
\label{eq:1-n}
||A||_1 = \max_j\sum_j|A(i,j)|
\end{equation}
where, $\zeta_j(A)$ denotes the $j$'th singular value of an arbitrary matrix $A$. Further, $\lambda_U$, $\lambda_P$, and $\lambda_Q$ are tunable parameters that are leveraged to balance out the two competing components of the objective functions: minimizing the rank of the recovered data versus the amount of changes made in the data during the recovery process. Mathematically, this means that by minimizing $f_U$, $f_P$, and $f_Q$, the data recovery process effectively minimizes the ranks of $M_U^*$, $M_P^*$, and $M_Q^*$. At the same time, the changes made in the data are kept small by penalizing the sparsity norm of matrices $\Delta S_U$, $\Delta S_P$, and $\Delta S_Q$. The three objectives $f_U$, $f_P$, and $f_Q$ are evaluated over the datasets that are generated by the same system (e.g., same feeder, lateral, or service transformer). However, these three datasets are not independent from each other due to the power flow constraints. Thus, the re-calibration of these three datasets cannot be performed separately using conventional low rank data recovery methods, such as robust PCA and PCP. To address this problem, we propose a multi-objective PCP-based model that can jointly refine three the SM datasets. The objective function minimizes the ranks of recovered data to realize the best achievable SM re-alignment. Moreover, to incorporate the inherent interdependencies of the three objectives, power flow equations are added as the constraints of the model. The proposed multi-objective optimization is as follows:

\begin{equation}
\label{eq:obj}
\min_{M_U^*,M_P^*,M_Q^*}\ \{f_U,f_P,f_Q\}
\end{equation}
\begin{equation}
\label{eq:constU}
s.t.\ \ ||M_U - M_U^* - \Delta S_U ||_F \leq \delta_U
\end{equation}
\begin{equation}
\label{eq:constP}
||M_P - M_P^* - \Delta S_P||_F \leq \delta_P
\end{equation}
\begin{equation}
\label{eq:constQ}
||M_Q - M_Q^* - \Delta S_Q ||_F \leq \delta_Q
\end{equation}
\begin{equation}
\label{eq:DistFlow}
M_U^* = M_P^* \cdot R^T + M_Q^* \cdot X^T + 1_{m \cross N}U_0
\end{equation}
where, $||\cdot||_F$ denotes the Frobenius norm of matrix, defined as follows:
\begin{equation}
\label{eq:frob}
||A||_F = \sqrt{\sum_i\sum_j A(i,j)}
\end{equation}
In addition, parameters $\delta_U$, $\delta_P$, and $\delta_Q$ are the standard deviations of the measurement/pseudo-measurement errors (obtained using knowledge of sensor tolerance or pseudo-measurement confidence intervals), matrices $R$ and $X$ represent the branch resistance and branch reactance of the network, respectively \cite{Qu2020}. $U_0$ is the primary voltage magnitude squared for the transformer to which the SMs are connected. The rationale behind constraints \eqref{eq:constU}, \eqref{eq:constP}, and \eqref{eq:constQ} is that the refined components (i.e., $M_U^*$, $M_P^*$, $M_Q^*$) are not exactly low rank and the asynchrony error components (i.e.,  $\Delta S_U$, $\Delta S_P$, $\Delta S_Q$) are not exactly sparse. Such soft constraints allow for slight deviations in the recovered data to compensate for SM measurement errors, which are consistent with our knowledge of measurement device confidence levels. Also, these allow utilities to minimize asynchrony error with noisy practical SM data, which particularly pertains to reactive power data that may be unavailable for residential customers. Constraint \eqref{eq:DistFlow} is obtained from the linear DistFlow in matrix form \cite{Baran1989}, which can enforce network physics and capture the inherent dependencies among datasets. The goal of this constraint is to ensure that the recovered SM data is feasible in power engineering context.

Our method follows the line of low rank data recovery techniques that have been commonly used in many areas \cite{PCAreview}. Unlike the black box methods that lack interpretability, the proposed model has a solid mathematical foundation to recover a low rank SM data matrix in the presence of gross asynchrony errors. Also, the dependencies among the datasets are basically encoded into the solution through a set of linear equality constraints. Such power flow models can be applied for arbitrary distribution systems. Note that the model is extendable to unbalanced systems in a straightforward way (i.e., full three-phase DistFlow is leveraged). Further, the proposed data recovery model makes no assumptions on system topology or load distribution, which ensures the performance of this model in other distribution systems.

\subsection{Solution Strategy}
A major challenge in solving the proposed data recovery model is the existence of power flow constraints \eqref{eq:DistFlow} that hinders the application of the existing closed-form dual solvers \cite{Zhou2010}. Another complication is that \eqref{eq:obj} has three non-smooth objective functions, which makes the problem non-differentiable. To efficiently tackle these challenges, we present a first-order Nesterov-like algorithm to solve the proposed multi-objective data recovery framework \cite{Nesterov2005}. The basic idea of our solution is to approximate the non-smooth objectives with differentiable surrogates. By applying this idea, the following surrogate components can be written for $f_U$ \cite{Aybat2011}:
\begin{equation}
    \label{eq:MUS}
    ||M_U^{*}||_{*} \approx \Gamma(M_U^{*},\mu_U) = \max_{||\alpha||_2 \leq 1} <M_U^{*},\alpha> - \frac{\mu_U}{2}||\alpha||^2_{F}
\end{equation}
\begin{equation}
    \label{eq:DSS}
    ||\Delta S_U||_{1} \approx \Psi(\Delta S_U,\nu_U) = \max_{||\beta||_{\infty} \leq 1} <\Delta S_U,\beta> - \frac{\nu_U}{2}||\beta||^2_{F}
\end{equation}
where, $\alpha$ and $\beta$ are auxiliary matrices, $\mu_U$ and $\nu_U$ are smoothness parameters, and $<\cdot,\cdot>$ is the Frobenius inner product \cite{Phillips1996}, calculated as follows:
\begin{equation}
    \label{eq:frobinner}
    <A,B> = \sum_i\sum_j A(i,j)\cdot B(i,j)
\end{equation}
Note that the non-differentiable norms are replaced with differentiable functions $\Gamma(\cdot,\cdot)$ and $\Psi(\cdot,\cdot)$ in \eqref{eq:MUS} and \eqref{eq:DSS}. The Lipschitz constants for the gradients of $\Gamma(\cdot,\cdot)$ and $\Psi(\cdot,\cdot)$ equal $\frac{1}{\mu_U}$ and $\frac{1}{\nu_U}$, respectively. Similar smooth surrogates are defined and calculated for the objectives $f_P$ and $f_Q$. By adopting this approximate alternative, the objectives in optimization \eqref{eq:obj} can be rewritten as a single-objective weighted averaging process by using a scalarization method \cite{MOO}. Since the relaxed problem is convex, the single-objective formulation is guaranteed to track all the Pareto-optimal solutions, given valid weight assignment to the objectives \cite{Deb2001}. The single-objective formulation can be rearranged as follows:
\begin{figure*}
\centering
\includegraphics[width=2\columnwidth]{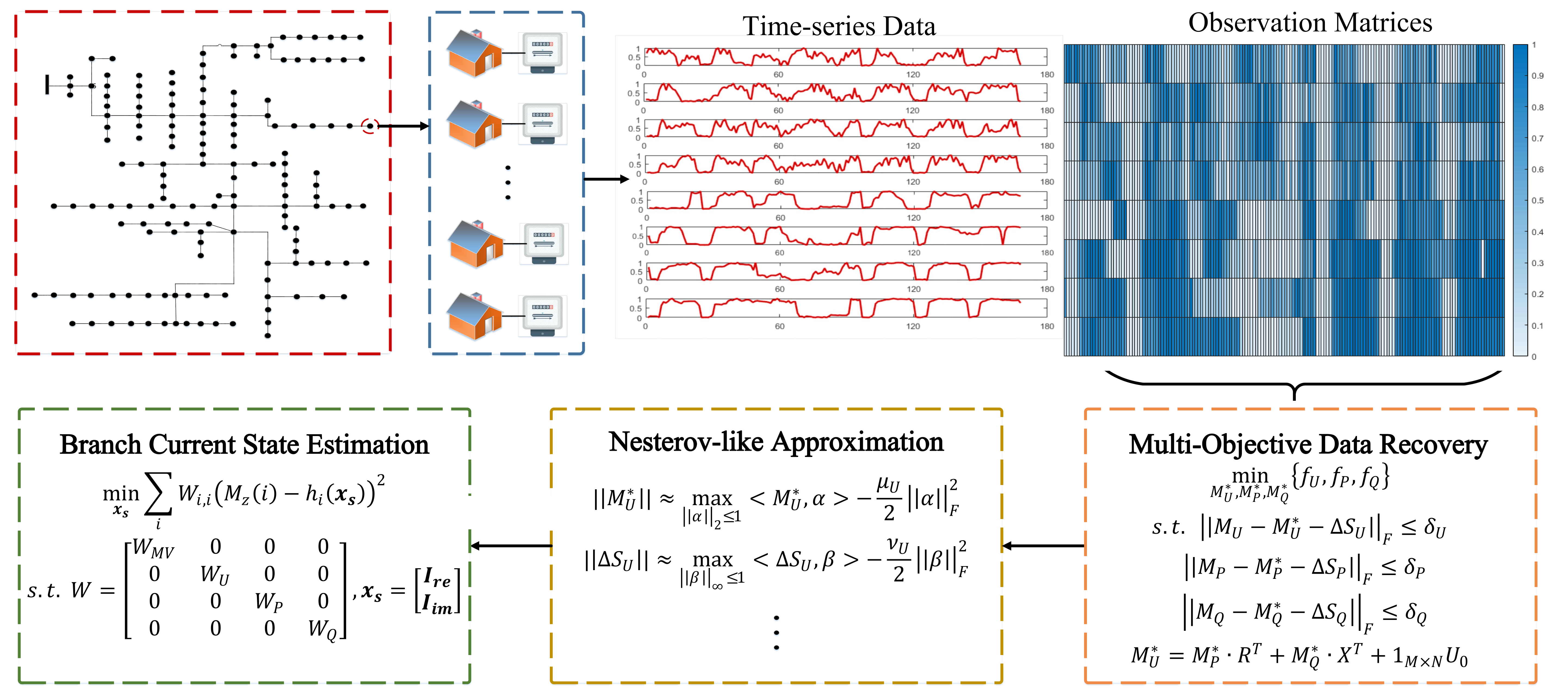}
\caption{Overall structure of the solution for grid monitoring.}
\label{fig:monitor}
\end{figure*}
\begin{equation}
    \label{eq:weightedsum}
    \min_{M_U^*,M_P^*,M_Q^*}\ \Gamma_T(M_U^*,M_P^*,M_Q^*) + \Psi_T(\Delta S_U,\Delta S_P,\Delta S_Q)
\end{equation}
\begin{equation}
    \label{eq:constraints1}
    s.t.\ \ \eqref{eq:constU}-\eqref{eq:DistFlow}
\end{equation}
here, the new objective function consists of two component: (I) $\Gamma_T$ quantifies the total approximate nuclear norm for all the SM datasets:
\begin{equation}
    \label{eq:GammaT}
    \begin{split}
        &\Gamma(M_U^*,M_P^*,M_Q^*) =\\ 
        &\omega_1\Gamma(M_U^{*},\mu_U) + \omega_2\Gamma(M_P^{*},\mu_P) + \omega_3\Gamma(M_Q^{*},\mu_Q)
    \end{split}
\end{equation}
where, $\omega_1$, $\omega_2$, and $\omega_3$ are the non-negative user-defined weights assigned to $f_U$, $f_P$, and $f_Q$, respectively. Assigning a larger weight to the objective function indicates that the function has a higher priority compared to a function with a smaller weight. Further, $ \omega_1 + \omega_2 + \omega_3 = 1$ needs to hold to ensure Pareto-optimality. (II) $\Gamma_T$ is the total approximate sparsity norm for all the SM datasets, as follows:
\begin{equation}
    \label{eq:PsiT}
    \begin{split}
        &\Psi(\Delta S_U,\Delta S_P,\Delta S_Q) =\\ 
        &\omega_1\lambda_U\Psi(\Delta S_U,\nu_U) + \omega_2\lambda_P\Psi(\Delta S_P,\nu_P) + \omega_3\lambda_Q\Psi(\Delta S_Q,\nu_Q)
    \end{split}
\end{equation}
This new data recovery formulation \eqref{eq:weightedsum} is both convex and differentiable. Given the new model, the Nesterov algorithm entails the following steps to solve SM data recovery problem:

\noindent\textbf{Step I -} \textit{Initialization}: $k\leftarrow 0$ (counter initialization); $M_U^*(0)\leftarrow M_U$, $M_P^*(0)\leftarrow M_P$, $M_Q^*(0) \leftarrow M_Q$, $\Delta S_U \leftarrow 0_{m\times N}$, $\Delta S_P\leftarrow 0_{m\times N}$, and $\Delta S_Q\leftarrow 0_{m\times N}$ (solution initialization).

\noindent\textbf{Step II -} \textit{Component-wise gradient calculation}: Obtain the gradients of components \eqref{eq:MUS} and \eqref{eq:DSS} for all the objective functions in the data recovery problem. As shown in \cite{Nesterov2005}, these gradients can be computed as follows:
\begin{equation}
    \label{eq:g_MUS}
    \nabla\Gamma(M_U^{*}(k),\mu_U) = \alpha^{*}(\mu_U)
\end{equation}
\begin{equation}
    \label{eq:g_DSS}
    \nabla\Psi(\Delta S_U(k),\nu_U) = \beta^{*}(\nu_U)
\end{equation}
where, $\alpha^{*}$ and $\beta^{*}$ are the optimal solutions of \eqref{eq:MUS} and \eqref{eq:DSS}, respectively, obtained for the latest values of $M_U^{*}$ and $\Delta S_U$ at iteration $k$. Similar gradient values can be obtained for surrogate components of active/reactive power data.

\noindent\textbf{Step III -} \textit{Aggregate gradient computation}: Insert the obtained gradients in Step II, to form the overall gradient values for the weighted averaging problem \eqref{eq:weightedsum}:
\begin{equation}
    \label{eq:g_w1}
        \nabla\Gamma_T(M_U^*,M_P^*,M_Q^*) =
        \omega_1\alpha^{*}(\mu_U) + \omega_2\alpha^{*}(\mu_P) + \omega_3\alpha^{*}(\mu_Q)
\end{equation}
\begin{equation}
    \label{eq:g_w2}
        \nabla\Psi_T(\Delta S_U,\Delta S_P,\Delta S_Q) = \omega_1\beta^{*}(\nu_U) + \omega_2\beta^{*}(\nu_P) + \omega_3\beta^{*}(\nu_Q)
\end{equation}

\noindent\textbf{Step IV -} \textit{Interim variable updates}: This step in the algorithm defines and updates several interim variables. These variables will be leveraged in the data refinement step. The idea is to apply gradient descent using the aggregate gradient components, obtained in Step III, while at the same time penalize deviations from the original measurements. Four interim matrices are defined: $Y_M$, $Y_S$, $Z_M$, and $Z_S$. While $Y_M$ and $Y_S$ are computed using the latest solution updates, on the other hand, $Z_M$ and $Z_S$ are obtained using the full history of the solution trajectory. Accordingly, the update process for $[Y_M,Y_S]$ is a convex and tractable optimization process, as follows:
\begin{equation}
    \label{eq:update1}
    \begin{split}
        &[Y_M,Y_S] = \argminA_{M,S} \{<\nabla\Gamma_T(M_U^*(k),M_P^*(k),M_Q^*(k)),M>\\
        & + <\nabla\Psi_T(\Delta S_U,\Delta S_P,\Delta S_Q),S> + \frac{L}{2}(||\Delta M ||_F^2 + ||\Delta S||_F^2)\}
    \end{split}
\end{equation}
\begin{equation}
    \label{eq:constraints2}
    s.t.\ \ \eqref{eq:constU}-\eqref{eq:DistFlow}
\end{equation}
where, $L$ is a weight parameter used for penalizing deviations from SM measurements. Here, the deviation from the original data are denoted as $\Delta M$ and $\Delta S$ (e.g., $\Delta M = M - [(M_U^*(0),M_P^*(0),M_Q^*(0)]$). Similarly, a convex optimization process is defined for updating $[Z_M,Z_S]$, considering full solver trajectory:
\begin{equation}
    \label{eq:update3}
        [Z_M,Z_S] = \argminA_{M,S} \{\tau(M,S) + \frac{L}{2}(||\Delta M ||_F^2 + ||\Delta S||_F^2)\}
\end{equation}
\begin{equation}
    \label{eq:constraints3}
    s.t.\ \ \eqref{eq:constU}-\eqref{eq:DistFlow}
\end{equation}
where, $\tau(M,S)$ is an average aggregate gradient factor with respect to solver history, defined as follows: 
\begin{equation}
    \label{eq:update2}
    \begin{split}
        \tau (M,S) =\sum_{i=0}^k <\nabla\Gamma_T(M_U^*(k),M_P^*(k),M_Q^*(k)),M> +\\ <\nabla\Psi_T(\Delta S_U,\Delta S_P,\Delta S_Q),S>
    \end{split}
\end{equation}

\noindent\textbf{Step V -} \textit{Data refinement}: Apply a weighted averaging process using the updated interim variables, from Step IV, to refine the SM data. Based on the suggestion in \cite{Aybat2011}, this weighted update process is written as follows:

\begin{equation}
    \label{eq:dataupdate1}
\left[
\begin{array}{c}
M_U^*(k+1)\\
M_P^*(k+1)\\
M_Q^*(k+1)
\end{array}
\right]\leftarrow (\frac{k+1}{k+3}) Y_M + (\frac{2}{k+3})Z_M
\end{equation}

\begin{equation}
    \label{eq:dataupdate2}
\left[
\begin{array}{c}
\Delta S_U(k+1)\\
\Delta S_P(k+1)\\
\Delta S_Q(k+1)
\end{array}
\right]\leftarrow (\frac{k+1}{k+3}) Y_S + (\frac{2}{k+3})Z_S
\end{equation}

\noindent\textbf{Step V-}\textit{Iterate and terminate}: $k\leftarrow k + 1$ and go to Step II until the maximum number of iterations is reached. Output the refined SM datasets, $M_U^*$, $M_P^*$, and $M_Q^*$, after algorithm convergence.

\section{Enhancing Grid Monitoring Robustness to SM Asynchrony Error}\label{sec:dsse}

Fig. \ref{fig:monitor} shows how our proposed data recovery technique can be integrated into grid monitoring systems as a pre-processor. The refined data is continuously fed to a branch current state estimation (BCSE) module to monitor the grid states in real-time, including the real and imaginary parts of currents of all branches \cite{Wang2004}. The BCSE method leverages a weighted least squares (WLS)-based solver to minimize the sum of squared residuals ($J$). This problem can be formulated as an optimization task over the distribution network given the recovered data samples $M_U^{*}$, $M_p^{*}$, $M_Q^{*}$ from our multi-objective PCP-based model, as follows:
\begin{equation}
\label{eq:bcse0}
\begin{split}
\min_{\pmb{x_s}} J &= \sum_{i} W_{i,i} (M_z(i) - h_i(\pmb{x_s}))^2\\
s.t. \ \ \ &M_z = \left[
\begin{array}{c}
M_{MV} \\
M_U^{*}(:)\\
M_P^{*}(:)\\
M_Q^{*}(:)\\
M_{PS}
\end{array}
\right]\\
W = &\left[
\begin{array}{ccccc}
W_{MV} & \pmb{0} & \pmb{0}& \pmb{0}& \pmb{0}\\
\pmb{0} & W_U &\pmb{0}& \pmb{0}& \pmb{0}\\
\pmb{0}& \pmb{0} &W_P& \pmb{0}& \pmb{0}\\
\pmb{0} & \pmb{0}& \pmb{0} & W_Q& \pmb{0}\\
\pmb{0} & \pmb{0}& \pmb{0} & \pmb{0}&  W_{PS}
\end{array}
\right]\\
&\pmb{x_s} = \left[
\begin{array}{c}
\pmb{I_{re}} \\
\pmb{I_{im}}
\end{array}
\right]
\end{split}
\end{equation}
where, $\pmb{x_s}$ is the grid state vector that contains current real/imaginary values for all the branches of the distribution system ($\pmb{I_{re}}/\pmb{I_{im}}$), and $M_z$ is the measurement vector. The measurement data includes the MV network synchronized sensor measurements ($M_{MV}$), including SCADA and $\mu$PMUs, if available, the refined SM data, $M_U^{*}$, $M_P^{*}$, $M_Q^{*}$, and the pseudo measurements $M_{PS}$ that can generated by our previous work \cite{yuanyx}. $h_i$ is the measurement function that maps state values to the $i$'th measurement variable, which is obtained based on the power flow equation. Furthermore, $W$ is a weight matrix that represents the solver's confidence level in each element of $M_z$. The matrix $W$ includes the measurement confidence weights, consisting of sub-matrces $W_{MV}$, $W_U$, $W_P$, $W_Q$, and $W_{PS}$ corresponding to $M_{MV}$, $M_U^{*}$, $M_P^{*}$, $M_Q^{*}$, and $M_{PS}$, respectively. These weight values are determined by the nominal accuracy levels of the senors as $W_{i,i} = \frac{1}{\sigma_i^2}$, where $\sigma_i^2$ is the $i$'th sensor error variance \cite{Baran1995}. The purpose of the weights is to devalue the importance of unreliable data sources in grid monitoring.

The WLS-based solution employs a gradient-based algorithm to find the optimal solutions for \eqref{eq:bcse0} (i.e., $\nabla_{\pmb{x_s}J = 0}$) \cite{RS2009}. The algorithm involves the following steps to estimate the states of the grid:

\noindent\textbf{Step I} - \textit{Receive input data}: Receive the recovered SM data, $M_U^{*}$, $M_P^{*}$, and $M_Q^{*}$ (see Section \ref{sec:MO}), and the latest measurement data from the primary network, $M_{MV}$. Concatenate the input data to form the measurement vector, $M_z$.

\noindent\textbf{Step II} - \textit{State initialization}: $k\leftarrow 0$; initialize the values of the states through randomization, $\pmb{x_s}(k)$ (to speed up the BCSE solver the values of states can be initialized using the solutions from the last time step.) 

\noindent\textbf{Step III} - \textit{Jacobian computation}: Update the \textit{Jacobian matrix}, $H$, using the gradients of the measurement function. The Jacobian captures the sensitivity of the measurements to the state variables:
\begin{equation}
\label{eq:jac}
H_{i,j} = \frac{\partial h_i(\pmb{x_s}(k))}{\partial\pmb{x_s}_j}
\end{equation}
The Jacobian matrix can be conveniently calculated for the BCSE method for feeders with known topology (e.g., see \cite{Wang2004} for details on how Jacobian can be obtained for various types of measurement functions.)

\noindent\textbf{Step IV} - \textit{Gain matrix computation}: Leverage the Jacobian matrix from Step III to obtain the gain matrix, $G$, as follows:

\begin{equation}
\label{eq:bcse3}
G(\pmb{x_s}(k)) = H^\top(\pmb{x_s}(k))WH(\pmb{x_s}(k))
\end{equation}

\noindent\textbf{Step V} - \textit{State update}: Update the values of the states using the gain matrix within the first order Newton-Gauss method, as follows: 

\begin{equation}
\label{eq:bcse4}
\pmb{x_s}(k + 1) \leftarrow \pmb{x_s}(k)+G^{-1}H^\top W (M_z - \pmb{h}(\pmb{x_s}(k)))
\end{equation}

\noindent\textbf{Step VI} - \textit{Iterate and terminate}: $k\leftarrow k + 1$; go back to Step III until convergence, i.e., $k \geq k_{max}$, with $ k_{max}$ being a user-defined maximum number of iteration for the BCSE algorithm.

\noindent\textbf{Step VII} - \textit{Roll the time window}: At the new time point, the data recovery is performed using the latest measurement data, according to \ref{sec:MO}. Go back to Step I.

\begin{figure*}[tbp]
      \centering
      \includegraphics[width=2\columnwidth]{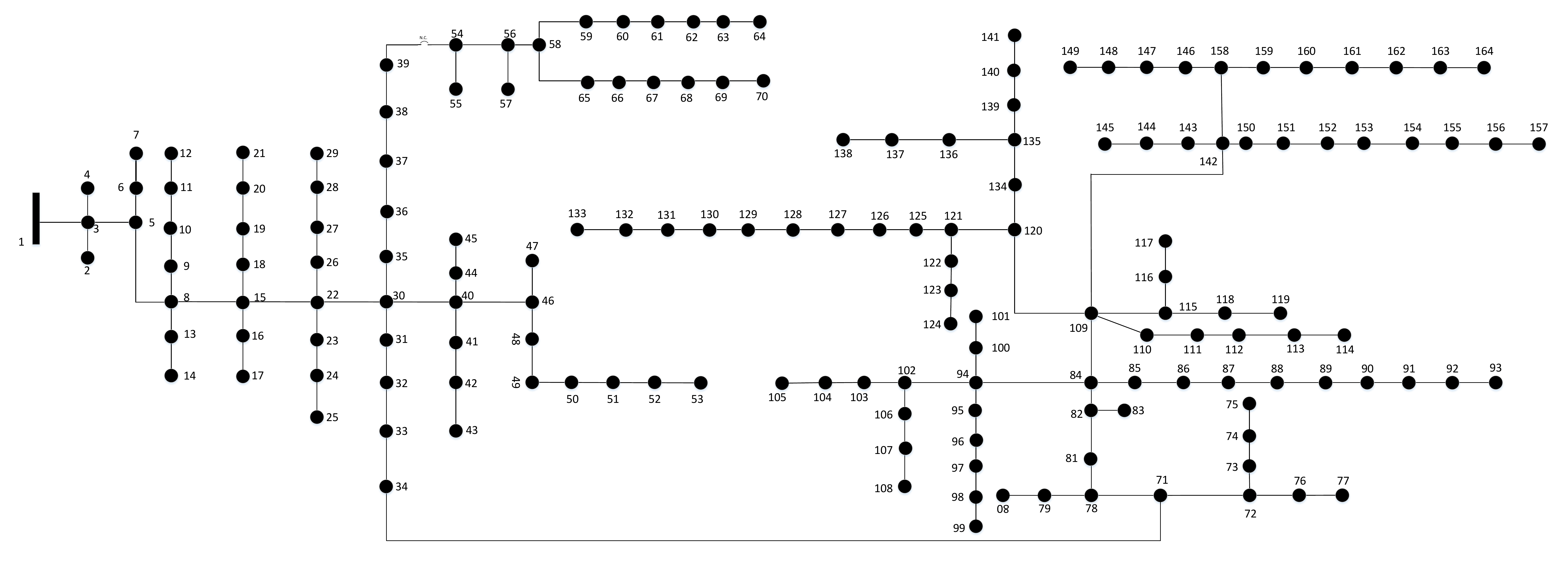}
\caption{164-node feeder topology.}
\label{fig:feeder}
\end{figure*}  

\section{Numerical Results}\label{sec:result}
\begin{figure}
\centering
\includegraphics[width=1\columnwidth]{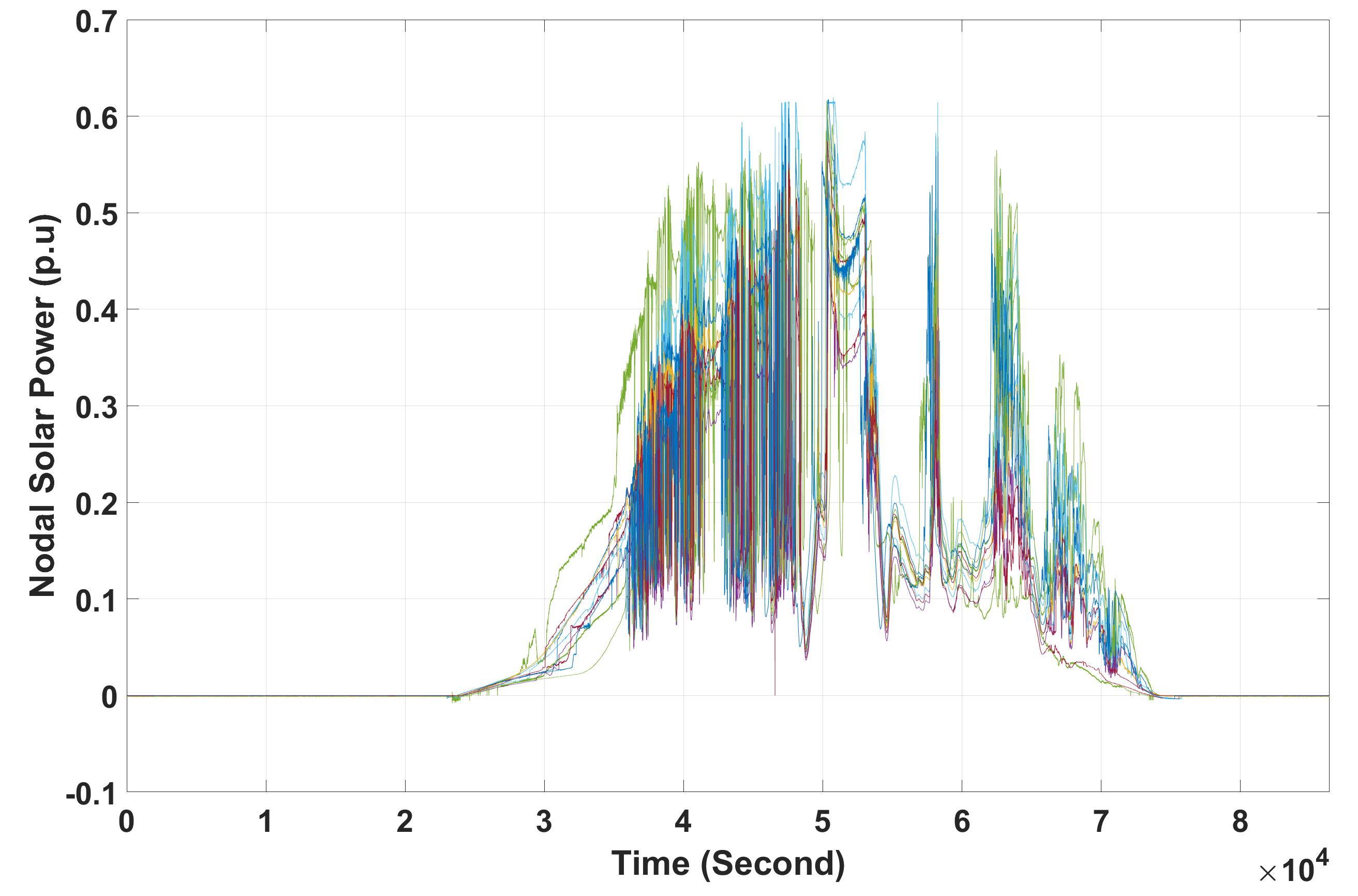}
\caption{PV generation data.}
\label{fig:pv}
\end{figure}

\begin{figure}
\centering
\includegraphics[width=1\columnwidth]{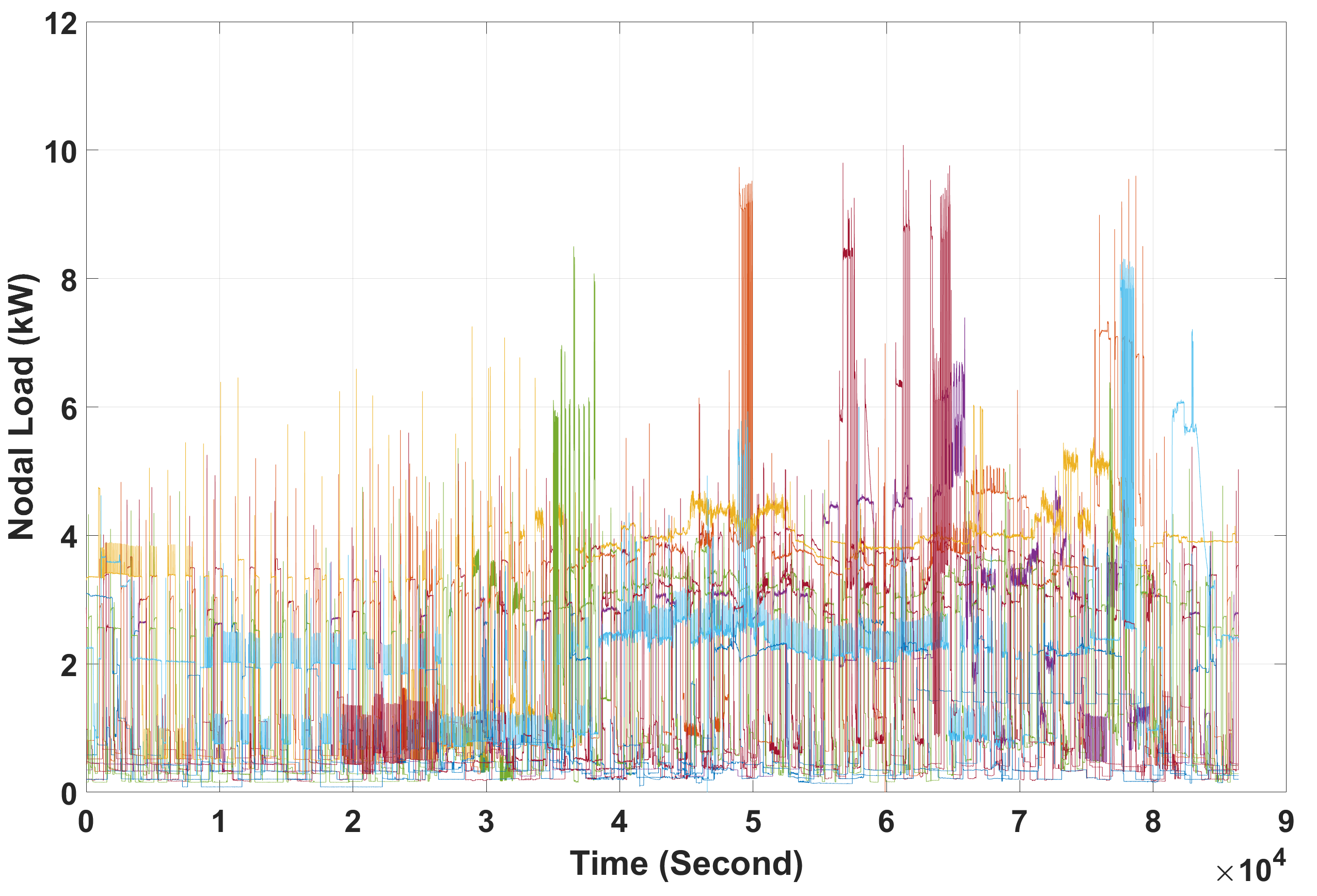}
\caption{Consumption data.}
\label{fig:load}
\end{figure}

The proposed data recovery and grid monitoring framework has been tested and validated using a fully observable feeder model shown in Fig. \ref{fig:feeder}. This feeder represents an unbalanced utility network in U.S. MidWest and consists of 164 nodes, which is publicly available online \cite{Test_system}. The details of the system model include network topology, line parameters, and standard electric components. The system has an average of 30\% solar-power-to-peak-load penetration level. The solar data is adopted from \cite{data}. The nodal time-series load demand is aggregated using a real-world 1-second-resolution household dataset and utilized as the input of the power flow analysis \cite{data}. The computed voltages are treated as the voltage measurements. The resolution of the SM measurements is 15-minute. To simulate realistic asynchronous SM measurements, we randomly sample the 1-second resolution data at 15-min rate at each node to represent SM measurements. Thus, in this work, the SM asynchrony strength of each customer can be anywhere between 0 to 900 s. Fig. \ref{fig:pv} and \ref{fig:load} show the original solar and load time-series data in a day at different nodes of the system. User-defined parameters within the proposed data recovery model, including coefficients of the optimization solver, have been tuned through try-outs over historical/simulation datasets. Basically, the values of these parameters are chosen when the residual of branch current state estimation is minimized. It should be noted that the high computational budget of this strategy does not impact the real-time performance of the proposed method since this parameter calibration is an offline process.

\begin{figure}
\centering
\includegraphics[width=1\columnwidth]{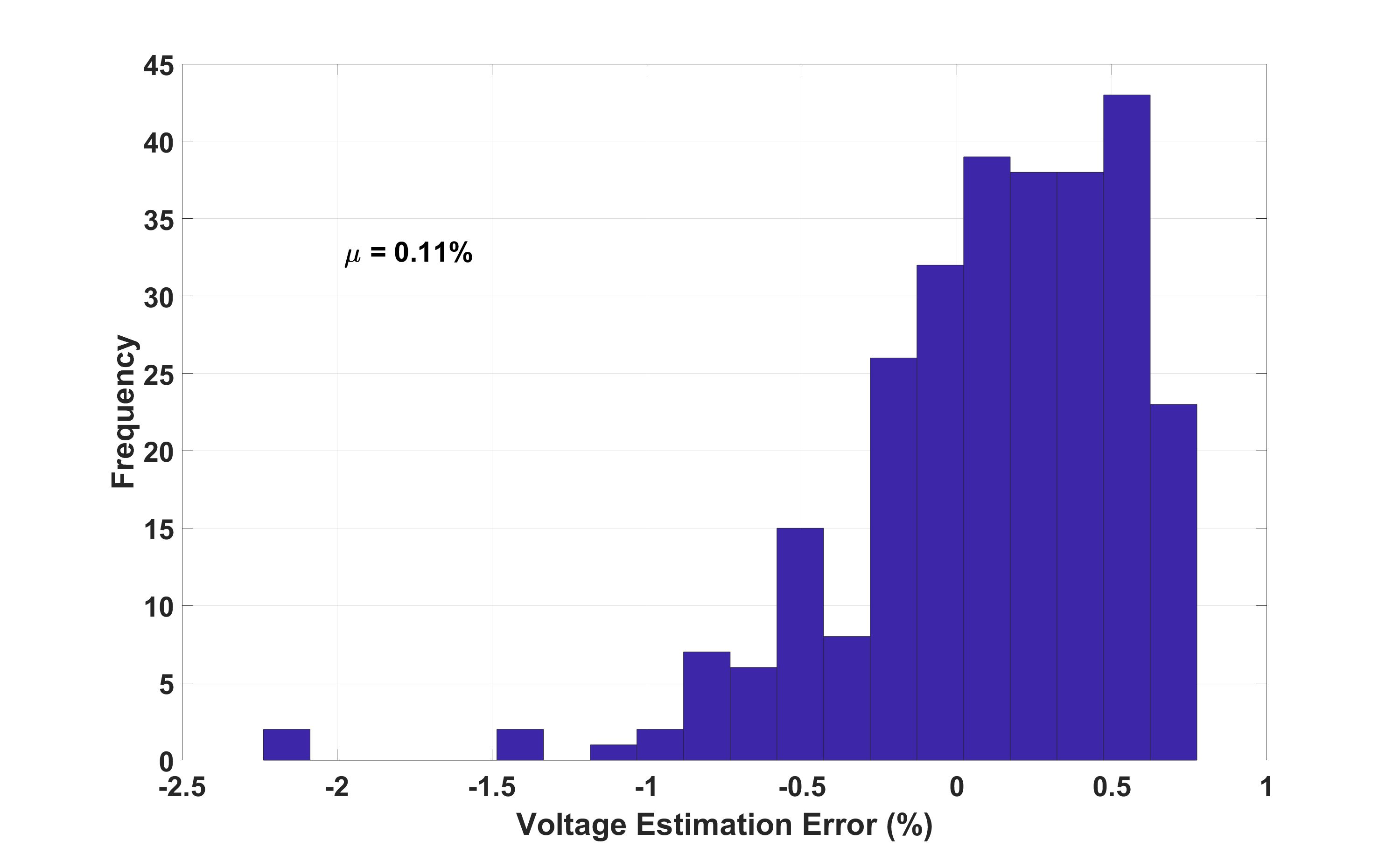}
\caption{Voltage recovery error.}
\label{fig:V_e}
\end{figure}

\begin{figure}
\centering
\includegraphics[width=1\columnwidth]{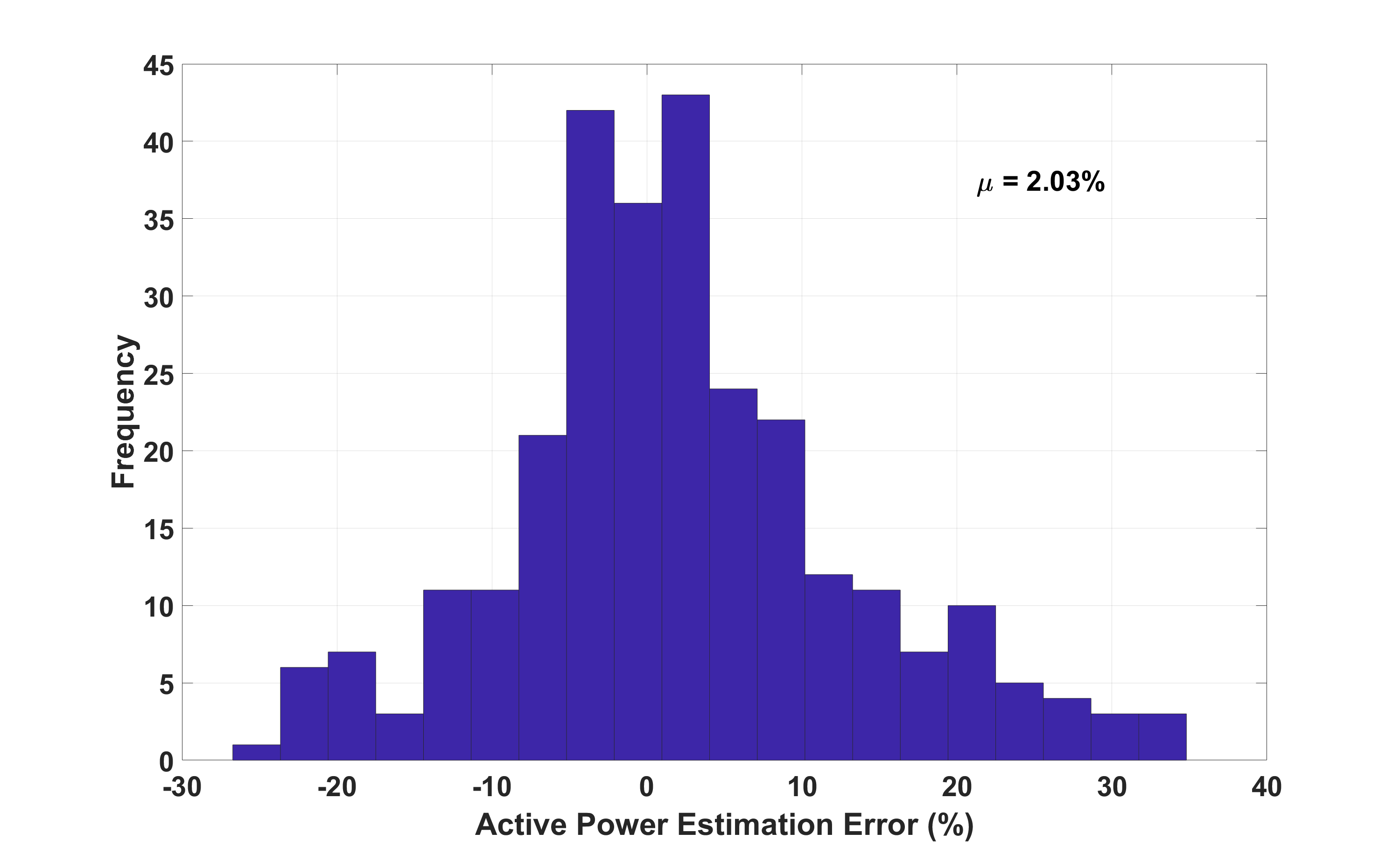}
\caption{Active power recovery error.}
\label{fig:P_e}
\end{figure}

\begin{figure}
\centering
\includegraphics[width=1\columnwidth]{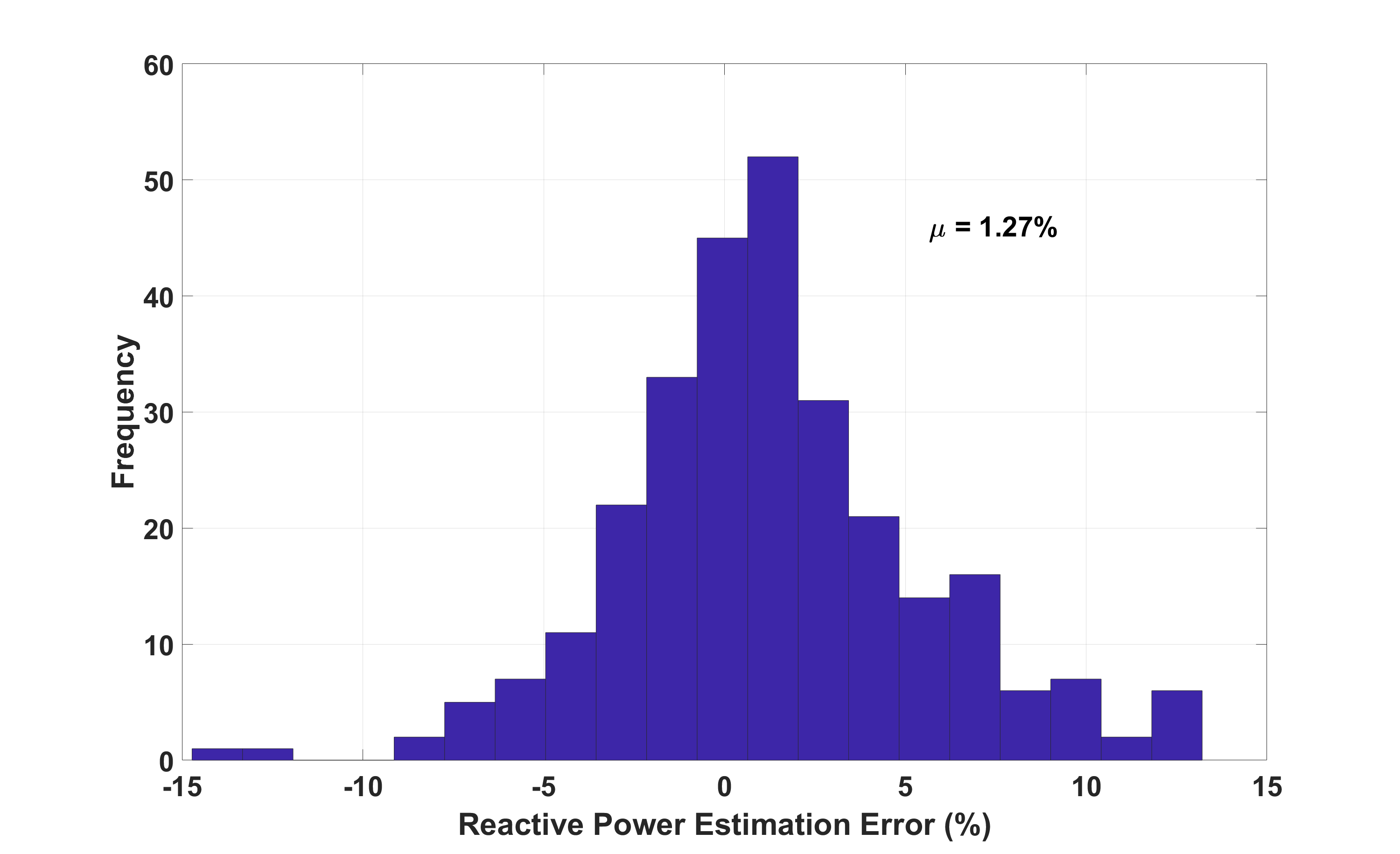}
\caption{Reactive power recovery error.}
\label{fig:Q_e}
\end{figure}

The case study is conducted on a standard PC with an Intel(R) Xeon(R) CPU running at 3.70 GHz and with 32.0 GB of RAM. Based on 500 Monte Carlo simulations, the average computational time is around 23 s, which is feasible in real-time applications. Fig. \ref{fig:V_e}, \ref{fig:P_e}, and \ref{fig:Q_e} show the average error histograms of the proposed data recovery method for voltage, active power, and reactive power, respectively. The error is calculated by comparing the actual values of various variables with the solutions of the recovery model. As can be observed, the recovery error values are maintained within low levels, which confirms the acceptable performance of the data recovery framework. Specifically, the mean average errors are 0.11\%, 2.03\%, and 1.27\% for voltage, active power, and reactive power, respectively. This also demonstrates that the proposed data refinement framework has the best performance over the SM voltage dataset, among the three datasets. This outcome is consistent with the correlation-driven nature of the data recovery model (i.e., nodal voltage measurements are highly correlated, which facilitates better refinement.)

\begin{figure}
\centering
\includegraphics[width=1\columnwidth]{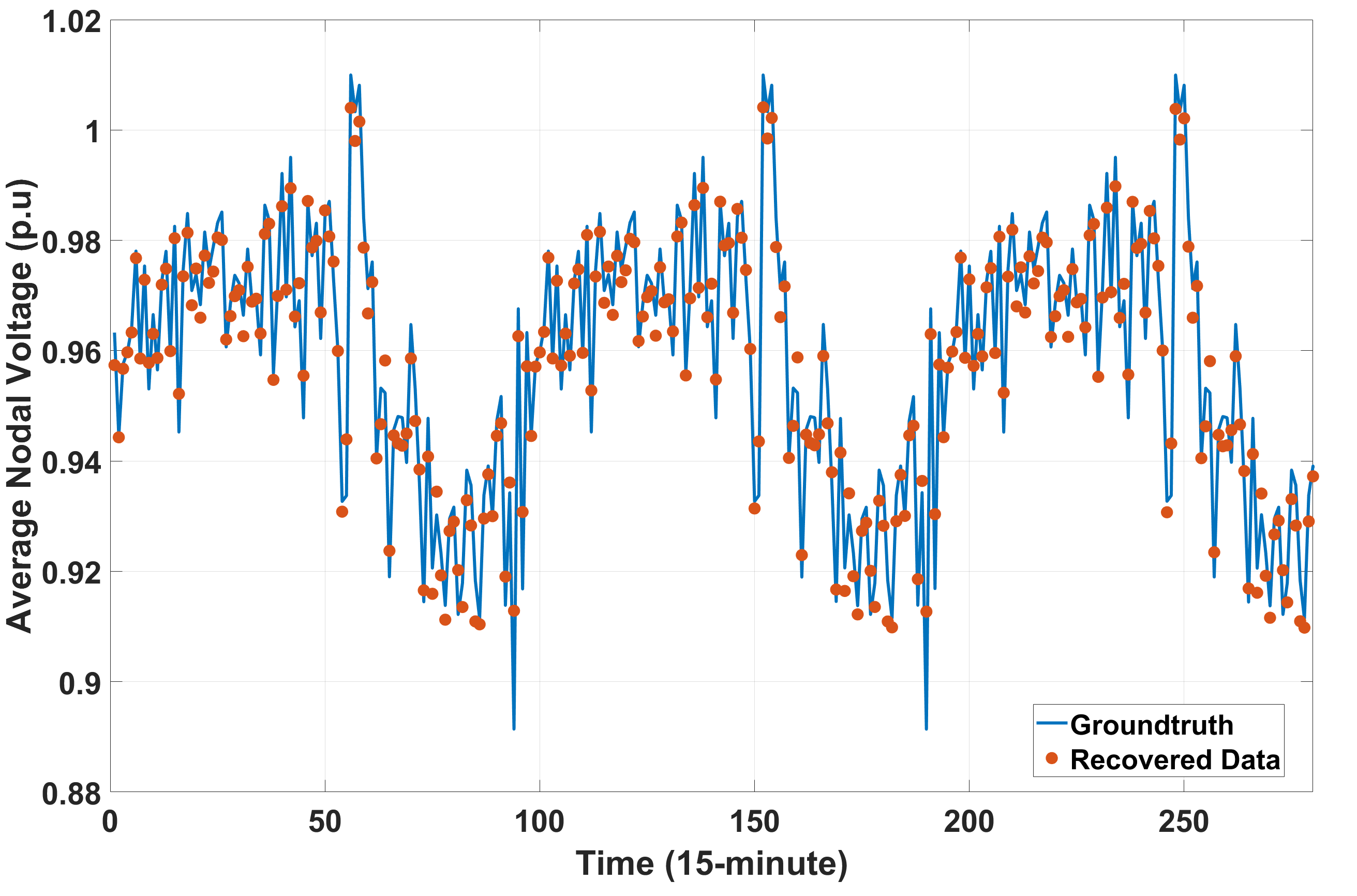}
\caption{Recovered average voltage data versus real (synchronized) time-series data.}
\label{fig:Vt}
\end{figure}

\begin{figure}
\centering
\includegraphics[width=1\columnwidth]{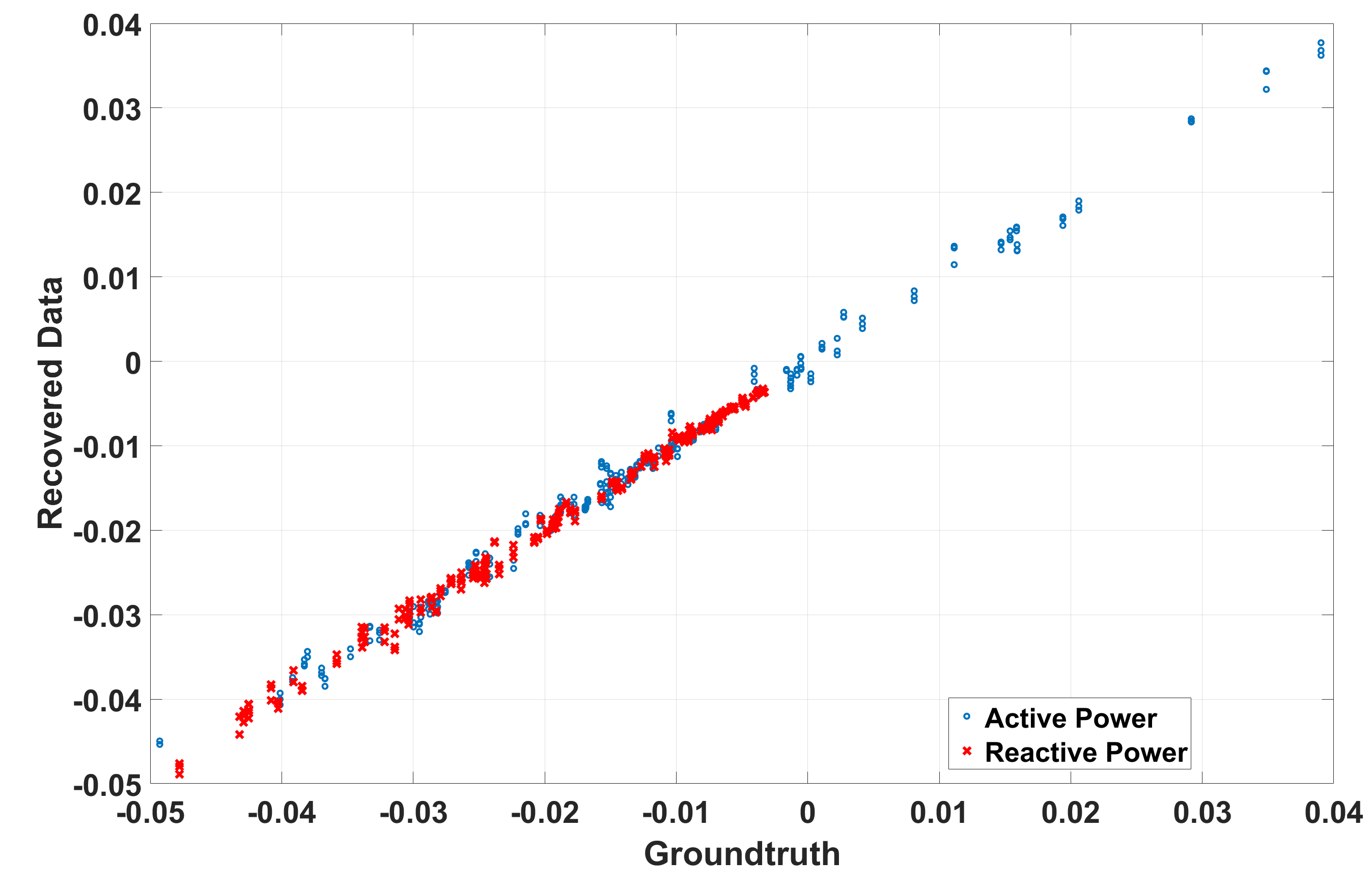}
\caption{Recovered nodal active/reactive power data versus real (synchronized) data.}
\label{fig:PQt}
\end{figure}

Fig. \ref{fig:Vt} compares the average value of recovered voltage data from the data refinement framework with the actual nodal voltage average (assuming synchronized sensors) within a sample time-window. As is observed in this figure, the developed algorithm closely follows the underlying signal. Fig. \ref{fig:PQt} shows a similar concept for active and reactive power datasets. As observed in this figure, the data recovery framework is basically an approximate identity mapping between the recovered data and the underlying (ideal) data. This corroborates the satisfactory performance of the model over real data in time domain. 

\begin{figure}
\centering
\subfloat[Without DistFlow\label{fig:woPF}]{
\includegraphics[width=0.5\textwidth]{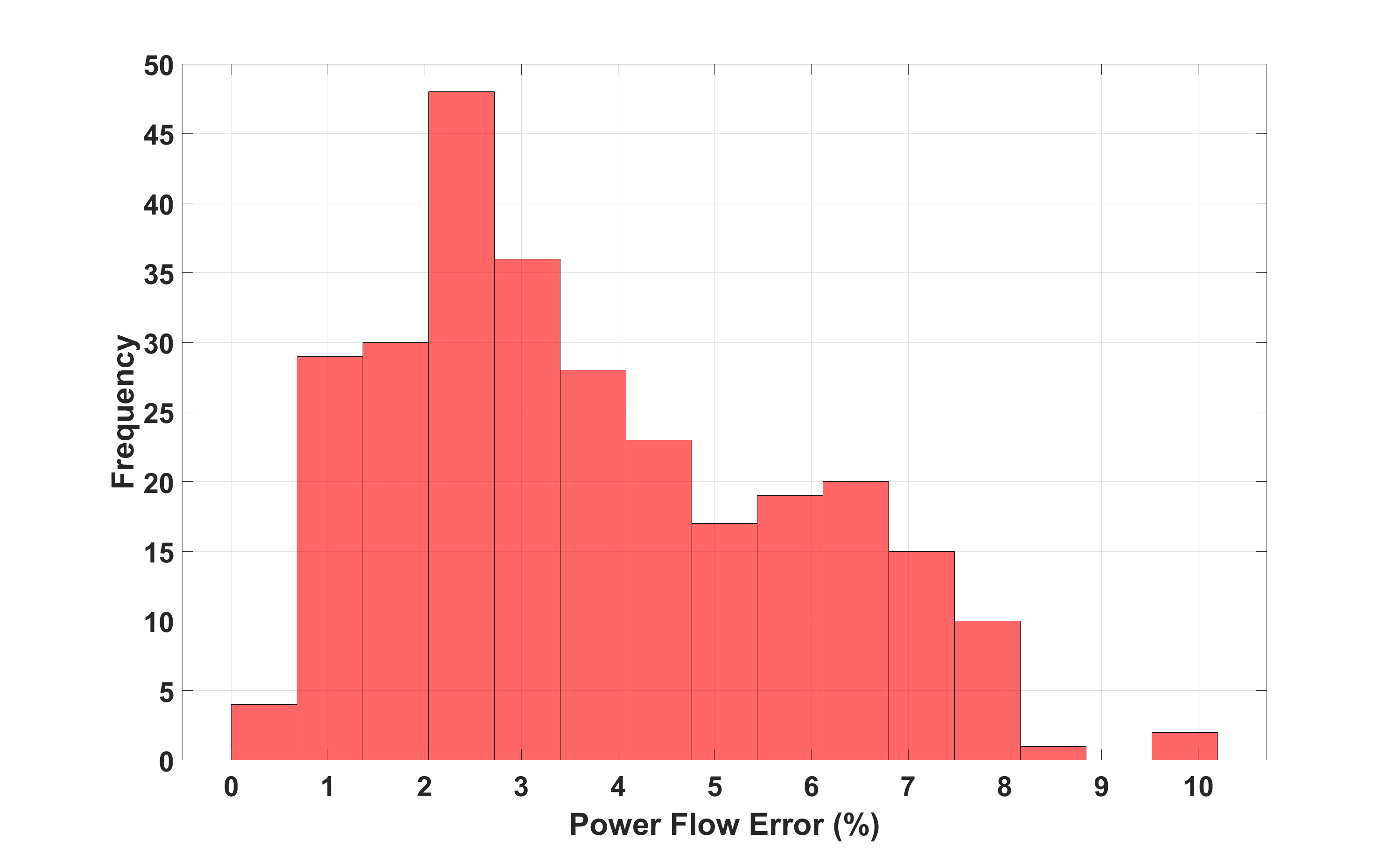}
}
\hfill
\subfloat[With DistFlow\label{fig:WPF}]{
\includegraphics[width=0.5\textwidth]{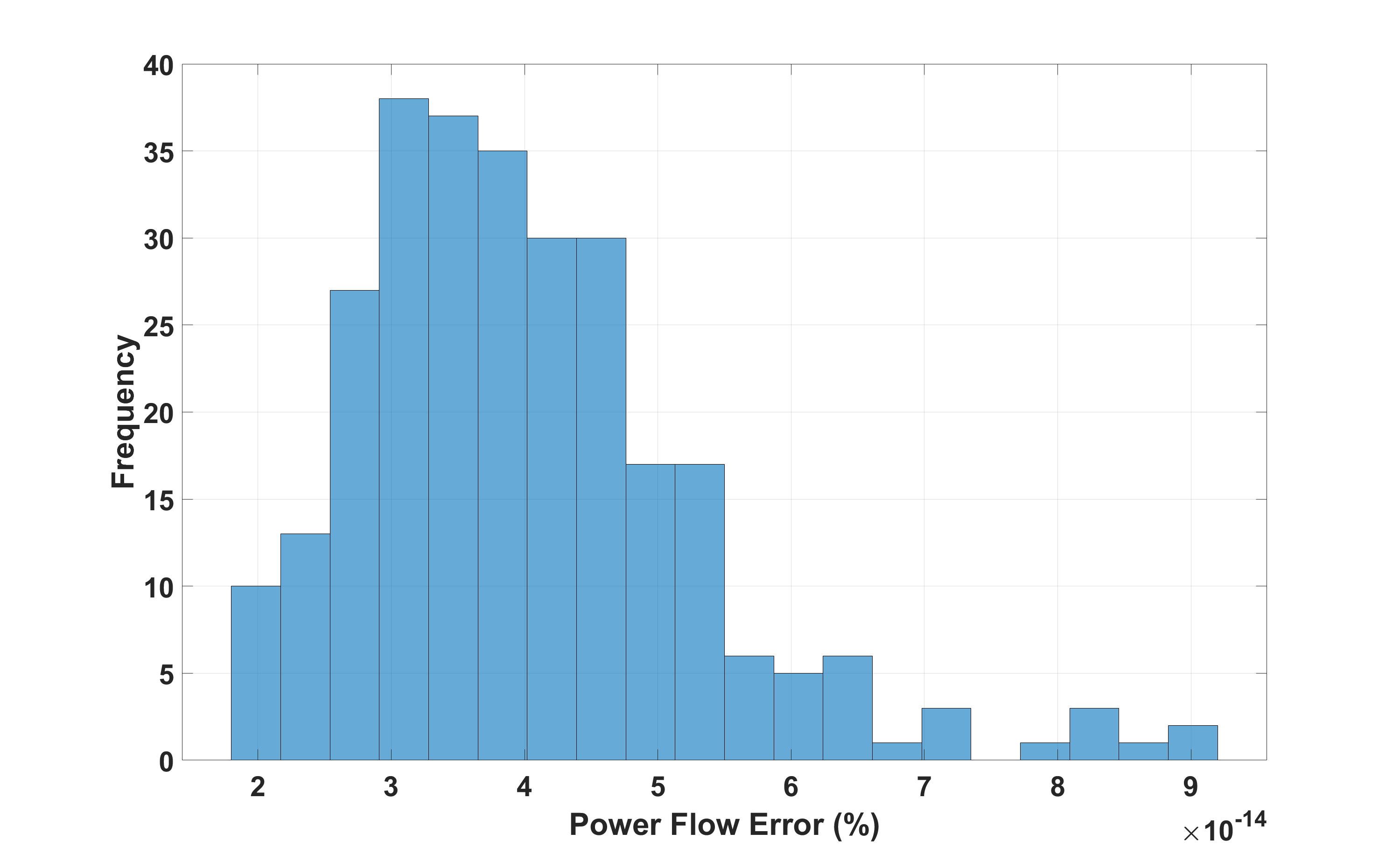}
}
\caption{Power flow error with and without DistFlow constraints.}
\label{fig:PF}
\end{figure}

Fig. \ref{fig:PF} shows the histogram of power flow error with and without leveraging the DisFlow equations within the proposed data recovery framework. As can be observed, having the DistFlow equations as constraints within the multi-objective data refinement model has resulted in a significant reduction in power flow errors. This demonstrates that the proposed method is able to output data that is consistent with network physics, while capturing the dependencies among all SM datasets.

\begin{figure}
\centering
\includegraphics[width=1\columnwidth]{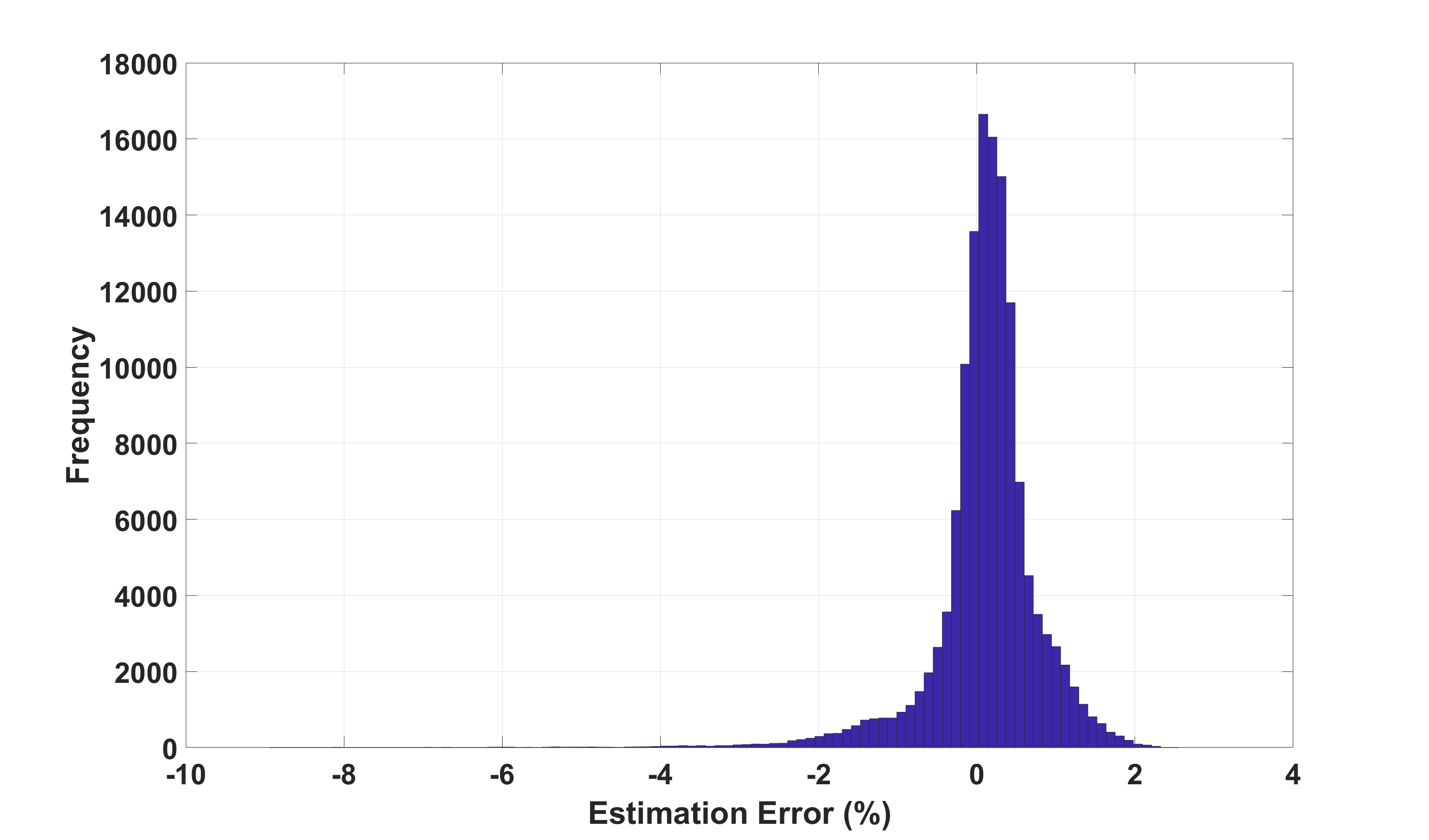}
\caption{BCSE error distribution.}
\label{fig:bcse_e}
\end{figure}

Finally, Fig. \ref{fig:bcse_e} depicts the histogram of system monitoring error after applying the data recovery framework. The mean percentage error (MPE) criterion is utilized to evaluate the performance of BCSE with our data recovery method, which is calculated by comparing the real state values ($\pmb{x_s}$), obtained from power flow simulations on the feeder model, with the estimated state values ($\hat{\pmb{x}}_{\pmb{s}}$), coming from the BCSE, as follows:
\begin{equation}
\label{eq:mape}
E = 100 \times \sum_{i}\frac{\pmb{x_s}(i) - \hat{\pmb{x}}_{\pmb{s}}(i)}{\pmb{x_s}(i)}
\end{equation}

As is observed in Fig. \ref{fig:bcse_e}, the DSSE error value is maintained at low levels, which demonstrates the successful integration of the data recovery solution into grid monitoring, which allows us to track the behavior of the feeder accurately. The mean estimation error value is 0.87\%.

\begin{figure}
\centering
\includegraphics[width=1\columnwidth]{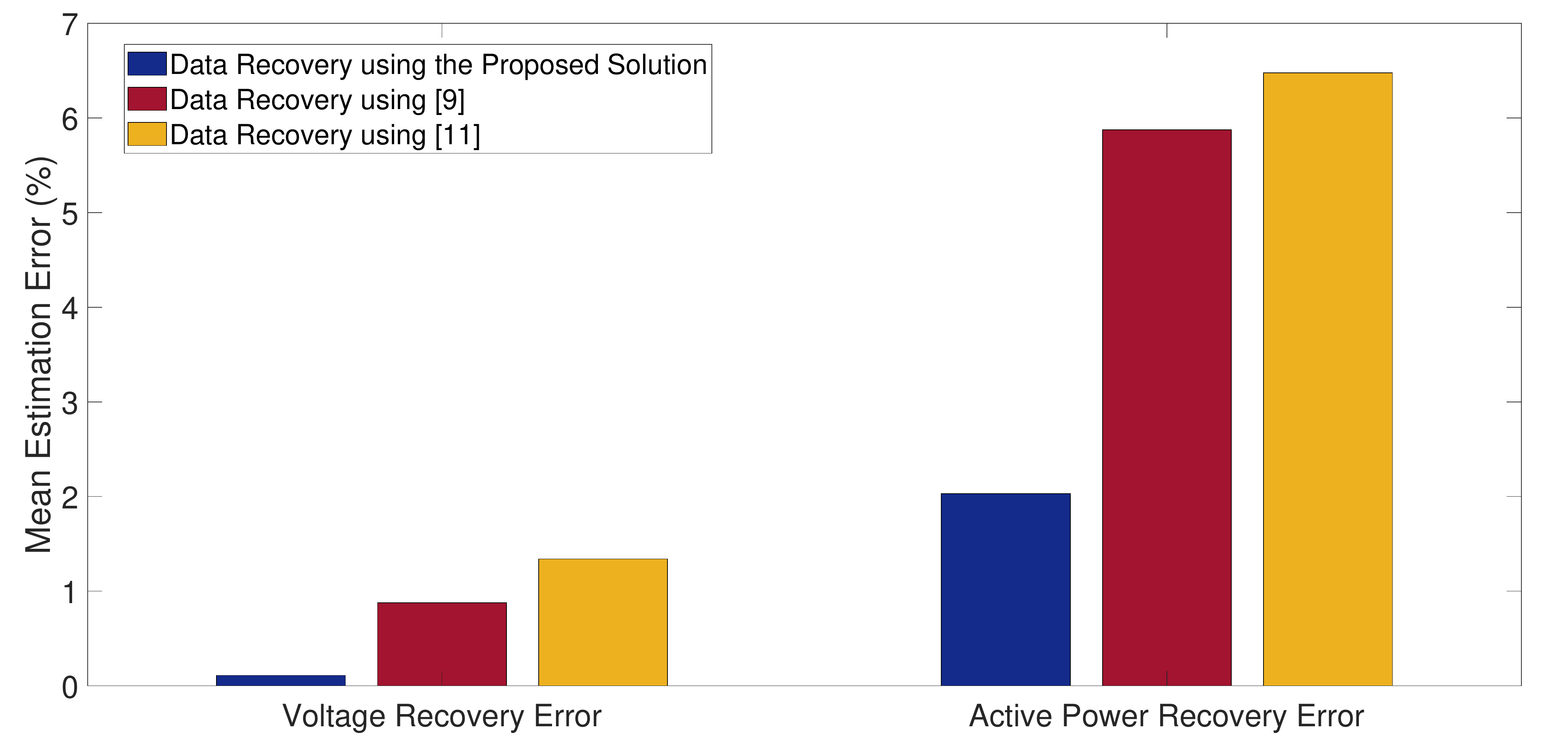}
\caption{Comparison results between \cite{Alimardani2015}, \cite{Bouwmans2014}, and the proposed method.}
\label{fig:bcompare}
\end{figure}

To further demonstrate the performance of the proposed SM data recovery method, We have conducted numerical comparisons with two previous methods, including a previous smart meter asynchrony mitigation method \cite{Alimardani2015} and a state-of-the-art low rank data recovery method \cite{Bouwmans2014}. The three methods are simulated with the same real-world datasets to calculate the accuracy of the methods. The comparison result is shown in \ref{fig:bcompare}. As demonstrated in the figure, in terms of voltage, the average recovery errors are 0.11\%, 0.877\%, and 1.34\% for the proposed solution, \cite{Alimardani2015} and \cite{Bouwmans2014}, respectively. In terms of active power, the average recovery errors are 2.03\%, 5.84\%, and 6.48\%, respectively. Hence, based on this dataset, the proposed method can achieve a better performance compared to the previous works.

\section{Conclusions}\label{sec:con}
In this paper, we have presented a multi-objective data recovery method to mitigate the impacts of SM asynchrony issues in distribution system real-time monitoring. The proposed method is able to refine voltage, active power, and reactive power datasets simultaneously within the same framework via a multi-objective formulation. The inherent dependencies among these measurements are captured by using DistFlow equations. Our solution considers both asynchrony errors and measurement errors, thus making the model more widely applicable to practical distribution systems. A first-order algorithm is presented to solve the proposed multi-objective data recovery model. This algorithm is based on Nesterov method for approximating non-differentiable optimization problems with smooth surrogates. To evaluate the proposed method, a real 164-node utility feeder with real data is utilized. The results show that SM asynchrony error mitigation is possible using the proposed method with good accuracy. In this work, the mean average data recovery error are about 0.11\%, 2.03\%, and 1.27\% for voltage magnitude, active power, and reactive power, respectively. Also, it can be observed that the DistFlow constraints can significantly reduce the inconsistency of recovered data with power flow equations. Based on the proposed data recovery method, the system state estimation error is less than 1$\%$.

\ifCLASSOPTIONcaptionsoff
  \newpage
\fi



\bibliographystyle{IEEEtran}
\bibliography{IEEEabrv,./bibtex/bib/main.bib}

\begin{thebibliography}{10}
\providecommand{\url}[1]{#1}
\csname url@samestyle\endcsname
\providecommand{\newblock}{\relax}
\providecommand{\bibinfo}[2]{#2}
\providecommand{\BIBentrySTDinterwordspacing}{\spaceskip=0pt\relax}
\providecommand{\BIBentryALTinterwordstretchfactor}{4}
\providecommand{\BIBentryALTinterwordspacing}{\spaceskip=\fontdimen2\font plus
\BIBentryALTinterwordstretchfactor\fontdimen3\font minus
  \fontdimen4\font\relax}
\providecommand{\BIBforeignlanguage}[2]{{%
\expandafter\ifx\csname l@#1\endcsname\relax
\typeout{** WARNING: IEEEtran.bst: No hyphenation pattern has been}%
\typeout{** loaded for the language `#1'. Using the pattern for}%
\typeout{** the default language instead.}%
\else
\language=\csname l@#1\endcsname
\fi
#2}}
\providecommand{\BIBdecl}{\relax}
\BIBdecl

\bibitem{EIA}
\BIBentryALTinterwordspacing
{Energy Information Administration}. (2020) Advanced metering count by
  technology type. [Online]. Available:
  \url{https://www.eia.gov/electricity/annual/html/epa_10_10.html}
\BIBentrySTDinterwordspacing

\bibitem{Lin2019}
J.~Lin, P.~Wang, S.~Guo, Y.~Shao, and Y.~Sheng, ``The credibility modelling and
  analysis of ami measurements for distribution system state estimation,''
  \emph{In {IEEE} Sustainable Power and Energy Conference (iSPEC)}, pp.
  1556--1560, 2019.

\bibitem{Primadianto2017}
A.~Primadianto and C.~N. Lu, ``A review on distribution system state
  estimation,'' \emph{{IEEE} Trans. Power Syst.}, vol.~32, no.~5, pp.
  3875--3883, Sep. 2017.

\bibitem{Deh2018}
K.~Dehghanpour, Z.~Wang, J.~Wang, Y.~Yuan, and F.~Bu, ``A survey on state
  estimation techniques and challenges in smart distribution systems,''
  \emph{{IEEE} Trans. Smart Grid}, vol.~10, no.~2, pp. 2312--2322, Sep. 2018.

\bibitem{Antonios2019}
I.~Antonios, H.~P. Schwefel, and L.~Lipsky, ``Approximation of the time
  alignment error for measurements in electricity grids,'' \emph{In 15th
  European Dependable Computing Conference (EDCC)}, pp. 153--158, 2019.

\bibitem{Schwefel2018}
H.~P. Schwefel, I.~Antonios, and L.~Lipsky, ``Impact of time interval alignment
  on data quality in electricity grids,'' \emph{In {IEEE} International
  Conference on Communications, Control, and Computing Technologies for Smart
  Grids (SmartGridComm)}, pp. 1--7, 2018.

\bibitem{Carvaro2019}
G.~Cavraro, E.~Dall'Anese, and A.~Bernstein, ``Dynamic power network state
  estimation with asynchronous measurements,'' \emph{National Renewable Energy
  Laboratory (NREL)}, 2019.

\bibitem{Bolognani2015}
S.~Bolognani, R.~Carli, and M.~Todescato, ``State estimation in power
  distribution networks with poorly synchronized measurements,'' \emph{In 53rd
  {IEEE} Conference on Decision and Control}, pp. 2579--2584, 2015.

\bibitem{Alimardani2015}
A.~Alimardani, F.~Therrien, D.~Atanackovic, J.~Jatskevich, and E.~Vaahedi,
  ``Distribution system state estimation based on nonsynchronized smart
  meters,'' \emph{{IEEE} Trans. Smart Grid}, vol.~6, no.~6, pp. 2919--2928,
  Jun. 2015.

\bibitem{Ni2016}
F.~Ni, P.~H. Nguyen, J.~F. Cobben, H.~E. van~den Brom, and D.~Zhao,
  ``Uncertainty analysis of aggregated smart meter data for state estimation,''
  \emph{In {IEEE} International Workshop on Applied Measurements for Power
  Systems (AMPS)}, pp. 1--6, 2016.

\bibitem{Bouwmans2014}
T.~Bouwmans and E.~H. Zahzah, ``Robust pca via principal component pursuit: A
  review for a comparative evaluation in video surveillance,'' \emph{Computer
  Vision and Image Understanding}, vol. 122, pp. 22--34, May 2014.

\bibitem{Rodriguez2013}
P.~Rodriguez and B.~Wohlberg, ``Fast principal component pursuit via
  alternating minimization,'' \emph{IEEE International Conference on Image
  Processing}, pp. 69--73, Sep. 2013.

\bibitem{Zhou2010}
Z.~Zhou, X.~Li, J.~Wright, E.~Candes, and Y.~Ma, ``Stable principal component
  pursuit,'' \emph{In {IEEE} International Symposium on Information Theory},
  pp. 1518--1522, 2010.

\bibitem{Gilbert1998}
G.~M. Gilbert, D.~E. Bouchard, and A.~Y. Chikhani, ``A comparison of load flow
  analysis using {DistFlow}, {Gauss-Seidel}, and optimal load flow
  algorithms,'' \emph{In IEEE Canadian Conference on Electrical and Computer
  Engineering (Cat. No. 98TH8341)}, vol.~2, pp. 850--853, May 1998.

\bibitem{Baran1989}
M.~Baran and F.~Wu, ``Optimal capacitor placement on radial distribution
  systems,'' \emph{IEEE Trans. Power Del.}, vol.~4, no.~1, pp. 725--734, Jan.
  1989.

\bibitem{Aybat2011}
N.~S. Aybat, D.~Goldfarb, and G.~Iyengar, ``Fast first-order methods for stable
  principal component pursuit,'' \emph{arXiv preprint arXiv:1105.2126}, vol.~4,
  no.~1, pp. 1--26, May 2011.

\bibitem{Phillips1996}
G.~M. Phillips and P.~J. Taylor, \emph{Theory and applications of numerical
  analysis}.\hskip 1em plus 0.5em minus 0.4em\relax Elsevier, 1996.

\bibitem{Qu2020}
G.~Qu and N.~Li, ``Optimal distributed feedback voltage control under limited
  reactive power,'' \emph{{IEEE} Trans. Power Syst.}, vol.~35, no.~1, pp.
  315--331, Jul. 2020.

\bibitem{PCAreview}
T.~Bouwmans and E.~H. Zahzah, ``Robust pca via principal component pursuit: A
  review for a comparative evaluation in video surveillance,'' \emph{Computer
  Vision and Image Understanding}, vol. 122, p. 1502242, 22-34 2014.

\bibitem{Nesterov2005}
Y.~Nesterov, ``Smooth minimization of non-smooth functions,''
  \emph{Mathematical Programming}, vol. 103, no.~1, pp. 127--152, May 2005.

\bibitem{MOO}
N.~Gunantara, ``A review of multi-objective optimization: Methods and its
  applications,'' \emph{Cogent Engineering}, vol.~5, no.~1, p. 1502242, 2018.

\bibitem{Deb2001}
K.~Deb, \emph{Multi-objective optimization using evolutionary
  algorithms}.\hskip 1em plus 0.5em minus 0.4em\relax West Sussex: John Wiely
  \& Sons, 2001.

\bibitem{Wang2004}
H.~Wang and N.~N. Schulz, ``A revised branch current-based distribution system
  state estimation algorithm and meter placement impact,'' \emph{IEEE Trans.
  Power Syst.}, vol.~19, no.~1, pp. 207--213, Feb 2004.

\bibitem{yuanyx}
Y.~{Yuan}, K.~{Dehghanpour}, F.~{Bu}, and Z.~{Wang}, ``A multi-timescale
  data-driven approach to enhance distribution system observability,''
  \emph{IEEE Transactions on Power Systems}, vol.~34, no.~4, pp. 3168--3177,
  2019.

\bibitem{Baran1995}
M.~E. Baran and A.~W. Kelley, ``A branch-current-based state estimation method
  for distribution systems,'' \emph{{IEEE} Trans. Power Syst.}, vol.~10, no.~1,
  pp. 483--491, Feb. 1995.

\bibitem{RS2009}
R.~{Singh}, B.~C. {Pal}, and R.~B. {Vinter}, ``Measurement placement in
  distribution system state estimation,'' \emph{IEEE Transactions on Power
  Systems}, vol.~24, no.~2, pp. 668--675, 2009.

\bibitem{Test_system}
F.~Bu, Y.~Yuan, Z.~Wang, K.~Dehghanpour, and A.~Kimber, ``A time-series
  distribution test system based on real utility datd,'' \emph{2019 North
  American Power Symposium (NAPS)}, pp. 1--6, 2019.

\bibitem{data}
\BIBentryALTinterwordspacing
C.~Holcomb, ``{Pecan Street Inc.}: A test-bed for {NILM},'' \emph{In
  International Workshop on Non-Intrusive Load Monitoring}, 2012. [Online].
  Available: \url{https://www.pecanstreet.org/}
\BIBentrySTDinterwordspacing

\end{thebibliography}
\end{document}